\documentclass[pra,aps,twocolumn,showpacs,preprintnumbers]{revtex4}
\usepackage{graphicx}
\usepackage{dcolumn}
\usepackage{bm}
\usepackage{amssymb}
\usepackage{epsfig}
\usepackage[english]{babel}
\usepackage{latexsym}
\usepackage{graphics}
\usepackage{subfigure}

\def\be{\begin{equation}}
\def\ee{\end{equation}}
\def\bea{\begin{eqnarray}}
\def\eea{\end{eqnarray}}
\def\bse{\begin{subequations}}
\def\ese{\end{subequations}}

\begin{document}

\title{Probing $n$-Spin Correlations in Optical Lattices}
\author{Chuanwei Zhang, V. W. Scarola, and S. Das Sarma }
\affiliation{Condensed Matter Theory Center, Department of Physics, University of
Maryland, College Park, MD 20742 }

\begin{abstract}
We propose a technique to measure multi-spin correlation functions of
arbitrary range as determined by the ground states of spinful cold atoms in
optical lattices. We show that an observation of the atomic version of the
Stokes parameters, using focused lasers and microwave pulsing, can be
related to $n$-spin correlators. We discuss the possibility of detecting not
only ground state static spin correlations, but also time-dependent spin
wave dynamics as a demonstrative example using our proposed technique.
\end{abstract}

\maketitle

\section{Introduction}

The advent of optical lattice confinement of ultracold atomic gases \cite%
{Oplat,Jaksch,Greiner,Paredes} opens the possibility of observing a vast
array of phenomena in quantum condensed systems \cite{Lewenstein}. In
particular, optical lattice systems may turn out to be the ideal tools for
the analog simulation of various strongly correlated interacting lattice
models (e.g. Hubbard model \cite{Jaksch,Greiner}, Kitaev model \cite{Kitaev}%
) studied in solid state physics. The great advantage of optical lattices as
analog simulators of strongly correlated Hamiltonians lies in the ability of
optical lattices to accurately implement lattice models without impurities,
defects, lattice phonons and other complications which can obscure the
observation of quantum degenerate phenomena in the solid state.

In this context optical lattices can support a variety of interacting spin
models which to date have been only approximately or indirectly observed in
nature or remain as rather deep but unobserved mathematical constructs.
Three exciting possibilities are currently the subject of active study \cite%
{Lewenstein}. The first (and the most direct) envisions simulation of
conventional condensed matter spin lattice models in optical lattices.
Quantum magnetism arising from strong correlation leads to many-body spin
ground states that can be characterized by spin order parameters. Spin order
can, in some cases, show long range behavior arising from spontaneous
symmetry breaking, e.g. ferromagnetism and antiferromagnetism. Such long
range spin ordering phenomena are reasonably well understood in most cases.
Recent work also relates conventional spin order parameters to entanglement
measures which yield scaling behavior near quantum phase transitions \cite%
{Entang1,Entang2}. The second possibility, simulation of topological spin
states, arises from the surprising fact that optical lattices can also (at
least in principle) host more complicated spin models previously thought to
be academic. The ground states of these models do not fall within the
conventional Landau paradigm, i.e. there is no spontaneously broken
symmetry, but show topological ordering and, as a result, display nontrivial
short range behavior in spin correlation functions. Examples include the
chiral spin liquid model \cite{Wen} and the Kitaev model \cite%
{Kitaev,Duan,Zoller,Zhang1}. And finally, optical lattices are also
particularly well suited to realize coherent and collective spin dynamics
because dissipation can be kept to suitably low levels \cite{Bloch}.

While optical lattices offer the possibility of realizing all three of the
above examples one glaring question remains. Once a suitable spin
Hamiltonian is realized, how do we observe the vast array of predicted
phenomena in spin-optical lattices? To date time of flight measurements have
proven to yield detailed information related to two types of important
correlation functions of many-body ground states of particles trapped in
optical lattices. The first, a first order correlation function (the
momentum distribution), indicates ordering in one-point correlation
functions \cite{Ketterle}. The second is a second order correlation function
(the noise distribution) which indicates ordering in two-point
density-density correlators \cite{Altman,Greiner2,Foelling,Spielman}. The
former can, for example, detect long range phase coherence while, as we will
see below, the latter is best suited to probe long range order in two-point
correlation functions, e.g. the lattice spin-spin correlation function. We
note that recent proposals suggest that time of flight imaging can in
principle be used to extract other correlation functions \cite{Duan2,Niu}.

In this paper we propose a technique to observe equal time $n$-spin
correlation functions characterizing \emph{both} long and short range spin
ordering useful in studying all three classes of spin lattice phenomena
mentioned above. Our proposal utilizes realistic experimental techniques
involving focused lasers, microwave pulsing and fluorescence detection to
effectively measure a general $n$-spin correlation function defined by: 
\begin{equation}
\xi \left\{ \alpha _{j_{k}},k=1,...,n\right\} \equiv \left\langle \Psi
\right\vert \prod_{k=1}^{n}\sigma _{j_{k}}^{\alpha _{j_{k}}}\left\vert \Psi
\right\rangle ,  \label{corfun}
\end{equation}%
where $\Psi $ is the many-body wavefunction of the atomic ensemble, $%
\{j_{k}\}$ is a set of sites, and $\sigma _{j_{k}}^{\alpha _{j_{k}}}$ ($%
\alpha _{j_{k}}=0,1,2,3$) are Pauli spin operators at sites $j_{k}$ with the
notation $\sigma ^{0}=I$, $\sigma ^{1}=\sigma ^{x}$, $\sigma ^{2}=\sigma
^{y} $, and $\sigma ^{3}=\sigma ^{z}$. Examples of order detectable with
one, two and three-spin correlation functions are magnetization $%
(\left\langle \sigma _{j}^{z}\right\rangle =1)$, anti-ferromagnetic order $%
(\left\langle \sigma _{j}^{z}\sigma _{j^{\prime }}^{z}\right\rangle
=(-1)^{j-j^{\prime }})$, and chiral spin liquid order ($\left\langle \sigma
_{j}\cdot \left( \sigma _{j^{\prime }}\times \sigma _{j^{\prime \prime
}}\right) \right\rangle =1$), to name a few.

In general our proposed technique can be used to experimentally characterize
a broad class of spin-lattice models of the form: 
\begin{equation}
H(J;{A})=J(t)\sum_{\{j_{k}\}}\left( A_{\{j_{k}\}}\prod_{k=1}^{M}\sigma
_{j_{k}}^{\alpha _{j_{k}}}\right) ,  \label{Hamiltonian}
\end{equation}%
where $J$ has dimensions of energy and can vary adiabatically as a function
of time, $t$, while the dimensionless parameters $A_{\{j_{k}\}}$ are kept
fixed. For example, $M=2$ represents the usual two-body Heisenberg model.
Several proposals now exist for simulating two-body Heisenberg models \cite%
{Lewenstein,Duan}. In the following we, as an example, consider optical
lattice implementations of the Heisenberg $XXZ$ model: 
\begin{equation}
H_{XXZ}(J;{\Delta })=J\left[ \sum_{\left\langle j,j^{\prime }\right\rangle
}\left( \sigma _{j}^{x}\sigma _{j^{\prime }}^{x}+\sigma _{j}^{y}\sigma
_{j^{\prime }}^{y}\right) +\Delta \sigma _{j}^{z}\sigma _{j^{\prime }}^{z}%
\right] ,  \label{Ham2}
\end{equation}%
where $\left\langle j,j^{\prime }\right\rangle $ denotes nearest neighbors
and $\Delta $ and $J$ are model parameters that can be adjusted by, for
example, varying the intensity of lattice laser beams \cite{Duan}.

The paper is organized as follows: in Section II we show that, in practice,
short range spin correlations are difficult to detect in noise correlation
measurements. Section III lays out a general procedure for detecting $n$%
-spin correlations with local probes. An experimental scheme for realizing
such a procedure is proposed and a quantitative feasibility analysis is
presented. Section IV is devoted to an investigation of the time-dependence
of correlation functions using the technique. In particular, we show how the
technique can be used to engineer and probe spin wave dynamics. We conclude
in Section V.

\section{Local Correlations in Time of Flight:}

We first discuss the measurement of spin-spin correlation functions by
analyzing spatial noise correlations (two point density-density
correlations, i.e. $\left( \left\langle n\left( \vec{r}\right) n\left( \vec{r%
}^{\prime }\right) \right\rangle -\left\langle n\left( \vec{r}\right)
\right\rangle \left\langle n\left( \vec{r}^{\prime }\right) \right\rangle
\right) /\left\langle n\left( \vec{r}\right) \right\rangle \left\langle
n\left( \vec{r}^{\prime }\right) \right\rangle $ in time of flight images
from atoms confined to an optical lattice modeled by the $XXZ$ Hamiltonian.
The ground states of this and a variety of spin models can be characterized
by the spin-spin correlation function between different sites. For instance,
the spin-spin correlation function in a one dimensional $XXZ$ spin chain
(with $J>0$), shows power-law decay 
\begin{equation}
\left\langle \sigma _{j}^{z}\sigma _{j^{\prime }}^{z}\right\rangle \sim
\left( -1\right) ^{j-j^{\prime }}/\left\vert j-j^{\prime }\right\vert ^{\eta
}  \label{spincor}
\end{equation}%
in the critical regime ($-1<\Delta \leq 1$), where $\eta =1/\left( 1-\frac{1%
}{\pi }\cos ^{-1}\Delta \right) $. In principle this correlation function
can be probed by spatial noise correlation in time of flight.

\begin{figure}[tbp]
\includegraphics[scale=0.5]{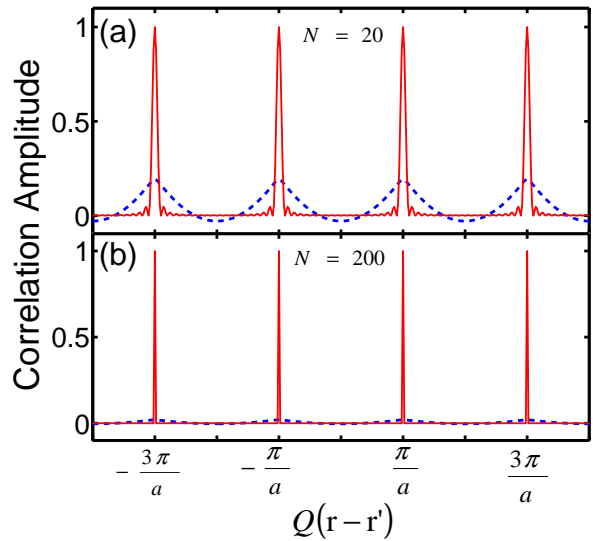}
\caption{ (Color online) Noise correlation plotted as a function of
wavevector of the one-dimensional $XXZ$ model. The solid (dashed) line
corresponds to a ground state with $\protect\eta =0$,$\,$long-range ($%
\protect\eta =2$, short-range) spin correlator. The amplitudes are
normalized by the maximum for anti-ferromagnetic order ($\protect\eta =0$).
The number of atoms in panel a (b) is $N=20$ ($N=200$).}
\label{fig:noise}
\end{figure}

We argue that, in practice, short range correlations (e.g. $\eta >1$ in the $%
XXZ$ model) are difficult to detect in time of flight noise correlation
measurements. To see this note that the noise correlation signal is
proportional to \cite{Altman}: 
\begin{equation}
G\left( Q\left( r-r^{\prime }\right) \right) =\sum_{j,j^{\prime }}e^{iQ\cdot
\left( j^{\prime }-j\right) a}\left\langle \sigma _{j}^{z}\sigma _{j^{\prime
}}^{z}\right\rangle ,  \label{noise}
\end{equation}%
where $Q$ is the lattice wave vector which gets mapped into coordinates $r$
and $r^{\prime }$ in time of flight on the detection screen, and $a$ is the
lattice spacing. Including normalization the noise correlation signal is
proportional to $N^{-1}$ for systems with long range order (e.g.
anti-ferromagnetic order giving $\eta =0$ in the above $XXZ$ model) but
shows a much weaker scaling for short range correlations. In fact the 
\textit{ratio} between correlators in a ground state with $\eta >1$ (short
range power law order) and $\eta =0$ (long range anti-ferromagnetic order)
scales as $N^{-1}$ making the state with power law correlations relatively
difficult to detect in large systems. To illustrate this we compare the
calculated noise correlation amplitude, $G$, in Fig.~\ref{fig:noise} for two
cases $\eta =0$ (solid line) and $\eta =2$ (dashed line) with $N=20$
(Fig.1a) and $N=200$ (Fig.1b) for the $1D$ $XXZ$ model. We see that the
correlation amplitude for short range (power-law) order is extremely small
in comparison to long range anti-ferromagnetic order for large $N$.


The small correlation signal originates from the fact that the noise
correlation method is in practice a conditional probability measuring
collective properties of the whole system, while short range spin
correlations describe local properties and are therefore best detected via
local operations. In the following we propose a local probe technique to
measure local correlations thus providing an experimental scheme which
compliments the time of flight$-$noise correlation technique, best suited
for detecting long range order.

\section{Detecting $n$-spin Correlation with Local Probes}

\subsection{General procedure}

We find that general $n$-spin correlators, $\xi \left\{ \alpha
_{j_{k}},k=1,...,n\right\} $, can be related to the Stokes parameters
broadly defined in terms of the local reduced density matrix $\rho =\text{Tr}%
_{\left\{ j_{k},k=1,...,n\right\} }\left\vert \Psi \right\rangle
\left\langle \Psi \right\vert $ on sites $\left\{ j_{k},k=1,...,n\right\} $,
where the trace is taken on all sites except the set $\left\{ j_{k}\right\} $%
. The Stokes parameters \cite{book} for the density matrix $\rho $ are $%
S_{\alpha _{j_{1}}....\alpha _{j_{n}}}=\text{Tr}\left( \rho
\prod_{k=1}^{n}\sigma _{j_{k}}^{\alpha _{j_{k}}}\right) $ leading to the
decomposition 
\begin{equation}
\rho =2^{-n}\sum_{\alpha _{j_{1}},...,\alpha _{j_{n}}=0}^{3}\left[ S_{\alpha
_{j_{1}}....\alpha _{j_{n}}}\prod_{k=1}^{n}\sigma _{j_{k}}^{\alpha _{j_{k}}}%
\right] .  \label{dens}
\end{equation}%
Using the theory of quantum state tomography \cite{book}, we find the $n$%
-spin correlators 
\begin{equation}
\xi \left\{ \alpha _{j_{k}},k=1,...,n\right\} =\prod_{k=1}^{n}\left(
P\left\{ \left\vert \phi _{\alpha _{j_{k}}}\right\rangle \right\} \pm
P\left\{ \left\vert \phi _{\alpha _{j_{k}}}^{\bot }\right\rangle \right\}
\right) ,  \label{corfun2}
\end{equation}%
where the plus (minus) sign indicates a $0$ (non-zero) index and $\left\{
\left\vert \phi _{\alpha _{j_{k}}}\right\rangle ,\left\vert \phi _{\alpha
_{j_{k}}}^{\bot }\right\rangle \right\} $ denote the measurement basis for
the atom at $j_{k}$. We define the measurement basis to be: $\left\vert \phi
_{1}\right\rangle =\left( \left\vert \downarrow \right\rangle +\left\vert
\uparrow \right\rangle \right) /\sqrt{2}$, $\left\vert \phi _{1}^{\bot
}\right\rangle =\left( \left\vert \downarrow \right\rangle -\left\vert
\uparrow \right\rangle \right) /\sqrt{2}$, $\left\vert \phi
_{2}\right\rangle =\left( \left\vert \downarrow \right\rangle +i\left\vert
\uparrow \right\rangle \right) /\sqrt{2}$, $\left\vert \phi _{2}^{\bot
}\right\rangle =\left( \left\vert \downarrow \right\rangle -i\left\vert
\uparrow \right\rangle \right) /\sqrt{2}$, $\left\vert \phi
_{3}\right\rangle =\left\vert \downarrow \right\rangle $, $\left\vert \phi
_{3}^{\bot }\right\rangle =\left\vert \uparrow \right\rangle $. Finally, $%
P\left\{ \left\vert \phi _{\alpha _{j_{k}}}\right\rangle \right\} $ is the
probability of finding an atom in the state $\left\vert \phi _{\alpha
_{j_{k}}}\right\rangle $.

The expansion of the product defining $\xi $ then yields a quantity central
to our proposal: 
\begin{equation}
\xi \left\{ \alpha _{j_{k}},k=1,...,n\right\} =\sum_{l=1}^{n}\left(
-1\right) ^{l}P_{l},  \label{measu}
\end{equation}%
where $P_{l}$ is the probability of finding $l$ sites in the states $%
\left\vert \phi _{j_{k}}^{\bot }\right\rangle $ and $n-l$ sites in $%
\left\vert \phi _{j_{k}}\right\rangle $. Eq.~(\ref{measu}) shows that the $n$%
-spin correlation function can be written in terms of experimental
observables. We can now write a specific example of the two-spin correlation
function (discussed in the previous section) in terms of observables: $\xi
\left\{ 3,3\right\} =P_{\left\vert \downarrow \right\rangle
_{j_{1}}\left\vert \downarrow \right\rangle _{j_{2}}}+P_{\left\vert \uparrow
\right\rangle _{j_{1}}\left\vert \uparrow \right\rangle _{j_{2}}}-\left(
P_{\left\vert \downarrow \right\rangle _{j_{1}}\left\vert \uparrow
\right\rangle _{j_{2}}}+P_{\left\vert \uparrow \right\rangle
_{j_{1}}\left\vert \downarrow \right\rangle _{j_{2}}}\right) $. In the
following subsection we discuss a specific experimental procedure designed
to extract precisely this quantity using local probes of cold atoms confined
to optical lattices.

\subsection{Proposed Experimental Realization}

We now describe and critically analyze an experimental procedure designed to
find the probabilities, $P_{l}$, from which we can determine the spin
correlation function $\xi \left\{ \alpha _{j_{k}},k=1,...,n\right\} $
through Eq. (\ref{measu}). To illustrate our technique we consider a
specific experimental system: $^{87}$Rb atoms confined on a single two
dimensional ($xy$ plane) optical lattice with two hyperfine ground states $%
\left\vert \downarrow \right\rangle \equiv \left\vert
F=1,m_{F}=-1\right\rangle $ and $\left\vert \uparrow \right\rangle \equiv
\left\vert F=2,m_{F}=-2\right\rangle $ chosen as the spin of each atom. Here
the atomic dynamics in the $z$ direction are frozen out by high frequency
optical traps \cite{Raizen}. However, the scheme can be directly applied to
three dimensional optical lattices by using one additional focused laser
which propagates along the $xy$ plane and plays the same role as the focused
laser (propagating along the $z$ axis) discussed in the following \emph{step
(II)}.

In the Mott insulator regime with one atom per lattice site spin
Hamiltonians, defined in Eq. (\ref{Hamiltonian}), may be implemented using
spin-dependent lattice potentials in the super-exchange limit \cite{Duan}.
The spin coupling $J\left( t\right) $, i.e., the overall prefactor in Eq. (%
\ref{Hamiltonian}), can be controlled by varying the lattice depth. Our
proposed experimental procedure will build on such spin systems, although it
can be generalized to other implementations where $H$ is generated by other
means. In the spin systems, the ground states are strongly correlated
many-body spin states and their properties can be characterized by the spin
correlations between atoms at different lattice sites.

\begin{figure}[t]
\includegraphics[scale=0.55]{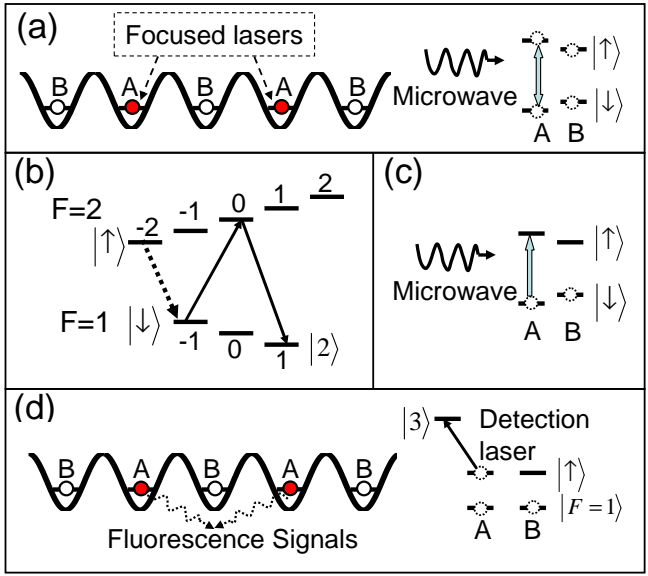}
\caption{(Color online) Schematic plot of the experimental procedure used to
measure $n$-spin correlation functions. $A$ indicates one of the target
atoms. $B$ indicates all other atoms. (a) Target atoms $A$ are transferred
to a suitable measurement basis using a combination of focused lasers and
microwave pulses (see step (II)). (b) Atoms at state $\left\vert \downarrow
\right\rangle $ are transferred to state $\left\vert 2\right\rangle $ by two
microwave $\protect\pi $-pulses, then atoms at state $\left\vert \uparrow
\right\rangle $ is transferred to state $\left\vert \downarrow \right\rangle 
$ by another microwave $\protect\pi $-pulse (see step (III). (c) Target
atoms $A$ are transferred back to state $\left\vert \uparrow \right\rangle $
from $\left\vert \downarrow \right\rangle $ (see step (IV)). (d) A detection
laser is applied to detect the probability of finding target atoms at $%
\left\vert \uparrow \right\rangle $ (see step (IV)). }
\label{rr2}
\end{figure}

\emph{Step (I)}: We start with a many-body spin state $\Phi _{0}$ and turn
off the spin-spin interactions generated by super exchange between lattice
sites. We achieve this by ramping up the spin-dependent lattice depth to $%
\sim 50E_{R}$ adiabatically with respect to the band splitting, where $%
E_{R}=h^{2}/2m\lambda ^{2}$ is the photon recoil energy and $\lambda $ is
the wavelength of the optical lattice. The ramp up time for each lattice is
chosen properly so that only the overall energy scale $J\left( t\right) $ in
the spin Hamiltonian (\ref{Hamiltonian}) is modified, which preserves the
highly correlated spin state $\Phi _{0}$. In the deep lattice, the time
scale for the spin-spin interactions ($\sim \hbar /J>10s$) becomes much
longer than the time ($\sim 1ms$) taken to perform the spin correlation
measurement. The spin-spin interactions play no role in the measurement
process and the following detection steps are quickly performed on this
\textquotedblleft frozen\textquotedblright\ many-body spin state $\Phi _{0}$.

\emph{Step (II)}: In this step, we selectively transfer target atoms $A$ at
site(s) $j_{k}$ (chosen aprior) to a suitable measurement basis $\left\{
\left\vert \phi _{j_{k}}\right\rangle ,\left\vert \phi _{j_{k}}^{\bot
}\right\rangle \right\} $ from initial states $\left\{ \left\vert \downarrow
\right\rangle _{j_{k}},\left\vert \uparrow \right\rangle _{j_{k}}\right\} $,
without affecting non-target atoms $B$ at other sites (Fig.2a). Selective
manipulation of quantum states of single atoms in optical lattices is
currently an outstanding challenge for investigating physics in optical
lattice, because spatial periods of typical optical lattices are shorter
than the optical resolution. In Ref. \cite{Zhang2}, a scheme for single atom
manipulation using microwave pulses and focused lasers is proposed and
analyzed in detail. Here we apply the scheme to selectively transfer atoms
between different measurement bases.

To manipulate a target atom $A$, we adiabatically turn on a focused laser
that propagates along the $\hat{z}$ axis having the maximal intensity
located at $A$. The spatially varying laser intensity $I\left( \mathbf{r}%
\right) $ induces position-dependent energy shifts%
\begin{equation}
\Delta E_{i}\left( \mathbf{r}\right) =\beta _{i}I\left( \mathbf{r}\right)
\label{eshift1}
\end{equation}%
for two spin states $\left\vert \downarrow \right\rangle $ and $\left\vert
\uparrow \right\rangle $, where the parameter $\beta _{i}$ for state $%
\left\vert i\right\rangle $ ($i=\downarrow $ or $\uparrow $) is determined
by the focused laser parameters. Different polarizations and detunings of
the focused laser lead to different $\beta _{i}$ and thus yield different
shifts of the hyperfine splittings $\left\vert \Delta E\left( \mathbf{r}%
\right) \right\vert =\left\vert \Delta E_{\uparrow }\left( \mathbf{r}\right)
-\Delta E_{\downarrow }\left( \mathbf{r}\right) \right\vert $ between two
spin states. We choose a $\sigma ^{+}$-polarized laser that drives the $%
5S\rightarrow 6P$ transition to obtain a small diffraction limit. The
wavelength $\lambda _{f}\approx 421nm$ (corresponding to a detuning $\Delta
_{0}=-2\pi \times 1209GHz$ from the $6^{2}P_{3/2}$ state) is optimized to
obtain the maximal ratio between energy splittings of two spin states and
the spontaneous scattering rate \cite{Zhang2}.

Because of the inhomogeneity of the focused laser intensity $I\left( \mathbf{%
r}\right) $, $\left\vert \Delta E\left( \mathbf{r}\right) \right\vert $
reaches a maximum at the target atom $A$ and decreases dramatically at
neighboring sites. Therefore the degeneracy of hyperfine splittings between
different atoms is lifted. By adjusting the focused laser intensity, the
differences of the energy shifts $\delta =\left\vert \Delta E\left( 0\right)
\right\vert -\left\vert \Delta E\left( \lambda /2\right) \right\vert $
between target atom $A$ and non-target neighboring atom $B$ can be varied
and calculated through Eq. (\ref{eshift1}). Here we choose $\delta =74E_{R}$
because it balances the speed and fidelity of single spin manipulation,
leading to less spontaneously scattered photons from the target atoms in the
spin-dependent focused lasers. To avoid excitations of atoms to higher bands
of the optical lattice, the rise speed of the focused laser intensity should
satisfy the adiabatic condition. We use the adiabatic condition to estimate
the ramp up time of the focused laser to be $35\mu s$ to give a $10^{-4}$
probability for excitation to higher bands.

We then change the measurement basis by applying a microwave $\pi /2$ pulse
that drives a suitable rotation to target atoms $A$ (Fig.2a). The microwave
is resonant with the hyperfine splitting between two spin states of the
target atoms $A$, but has a detuning larger than $\delta $ for non-target
atoms $B$. Consider a pulse with Rabi frequency $\Omega \left( t\right)
=\Omega _{0}\exp \left( -\omega _{0}^{2}t^{2}\right) $ ($-t_{f}\leq t\leq
t_{f}$) and parameters $\omega _{0}=14.8E_{R}/\hbar $, $\Omega
_{0}=13.1E_{R}/\hbar $ and $t_{f}=5/\omega _{0}$. The pulse transfers the
measurement basis of the target atoms $A$ in $16.9\mu s$, while the change
in the quantum state of non-target atoms is found to be below $3\times
10^{-4}$ by numerically integrating the Rabi equation that describes the
coupling between two spin states by the microwave pulse. The focused lasers
are adiabatically turned off after the microwave pulse. During the step
(II), the probability for spontaneous scattering of one photon from target
atoms inside the focused laser can be roughly estimated as $P\approx
\int_{\tau _{i}}^{\tau _{f}}\frac{\Gamma }{\hbar \vartheta }V\left( t\right)
dt\sim 2\times 10^{-4}$, where $\tau _{i}$ and $\tau _{f}$ represent the
times when the focused laser is turned on and off, $\Gamma $ is the decay
rate of $6P$ state, $\vartheta $ is the detuning, and $V\left( t\right) $ is
the potential depth of the focused laser.

The distance between any two target atoms can be as short as a lattice
spacing. This is because the basis transfer processes for different target
atoms are preformed sequentially in time, i.e., the process for an atom at
site $j_{2}$ starts after the process for an atom at site $j_{1}$ is
accomplished. In the special case that sites $j_{1}$ and $j_{2}$ are
spatially well separated and the final basis $\left\{ \left\vert \phi
_{j_{k}}\right\rangle ,\left\vert \phi _{j_{k}}^{\bot }\right\rangle
\right\} $ at two sites are the same, the transfer process can be done
simultaneously for two sites with one microwave pulse.

\emph{Step (III):} In this step, we transfer all atoms to the $\left\vert
F=1\right\rangle $ hyperfine level (Fig. 2b) to avoid stray signal in the
detection step (IV). We apply two $\pi $ microwave pulses to transfer all
atoms at $\left\vert \downarrow \right\rangle $ first to $\left\vert
F=2,m_{F}=0\right\rangle $ then to $|2\rangle \equiv \left\vert
F=1,m_{F}=1\right\rangle $. Another $\pi $ microwave pulse is then applied
to transfer all atoms at $\left\vert \uparrow \right\rangle $ to $\left\vert
\downarrow \right\rangle $. The $\pi $ microwave pulse can be implemented
within $12.5\mu s$ for a microwave Rabi frequency $\Omega =2\pi \times 40$%
kHz.

\emph{Step (IV):} We transfer target atoms $A$ at $j_{k}$ from $\left\vert
\downarrow \right\rangle $ back to $\left\vert \uparrow \right\rangle $ with
the assistance of the focused lasers and microwave pulses (Fig.2c), using
the same atom manipulation procedure as that described in step (II). We then
apply a detection laser resonant with $\left\vert \uparrow \right\rangle
\rightarrow \left\vert 3\right\rangle \equiv \left\vert
5^{2}P_{3/2}:F=3,m=-3\right\rangle $ to detect the probability of finding
all target atoms at $\left\vert \uparrow \right\rangle $ (corresponding to
the basis state $\left\vert \phi _{j_{k}}^{\bot }\right\rangle $ because we
transferred atoms to the measurement basis in step (II)) (Fig. 2d). The beam
size of the detection laser should be large enough so that target atoms at
different sites $\left\{ j_{k}\right\} $ experience the same laser
intensity, and scatter the same number of photons if they are in the $%
\left\vert \uparrow \right\rangle $ state. The fluorescence signal (the
number of scattered photons) is from one of the $n+1$ quantized levels,
where the $l$-th level ($l=0,...,n$) corresponds to states with $l$ sites of
target atoms on state $\left\vert \uparrow \right\rangle $ ($\left\vert \phi
_{j_{k}}^{\bot }\right\rangle $). By repeating the whole process many times,
we obtain the probability distribution $P_{l}$, and thus the spin
correlation function $\xi \left\{ \alpha _{j_{k}},k=1,...,n\right\} $ via
Eq.~(\ref{measu}). 

\begin{figure}[b]
\includegraphics[scale=0.43]{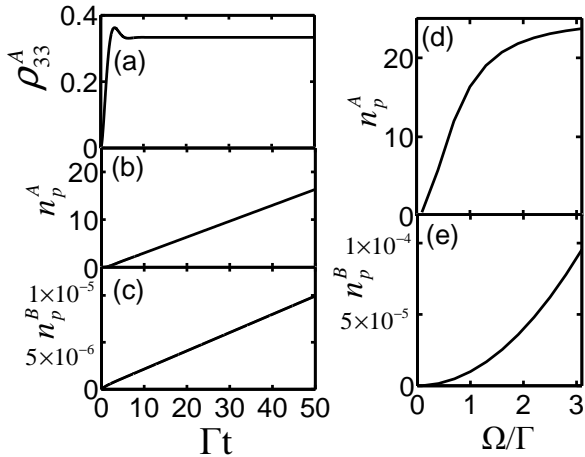}
\caption{$(a)$ Time evolution of the probability for the target atoms $A$ to
be in the excited state $\left\vert 3\right\rangle $. $(b)$ The number of
scattering photons $n_{p}^{A}$ versus time for atoms $A$ at state $%
\left\vert \uparrow \right\rangle $. $(c)$ The number of scattering photons $%
n_{p}^{B}\,$\ versus time for atoms ($A$ or $B$) at states $\left\vert
\downarrow \right\rangle $ and $\left\vert 2\right\rangle $. $(d)$ The
number of scattering photons $n_{p}^{A}$ versus the Rabi frequency of the
resonant laser for atoms $A$ at state $\left\vert \uparrow \right\rangle $. $%
(e)$ The number of scattering photons $n_{p}^{B}$ versus the Rabi frequency
of the resonant laser for atoms ($A$ or $B$) at states $\left\vert
\downarrow \right\rangle $ and $\left\vert 2\right\rangle $. The time period
for $(d)$ and $(e)$ is $50/\Gamma $.}
\label{fig:error}
\end{figure}

The scattering photons of the fluorescence signals come mostly from the
target atoms $A$ at state $\left\vert \uparrow \right\rangle $ (Fig.3).
Signal from atoms at any $\left\vert F=1\right\rangle $ state is suppressed
because of the large hyperfine splitting ($\nu \approx 2\pi \times 6.8GHz$)
between $\left\vert F=1\right\rangle $ and $\left\vert F=2\right\rangle $
states. Atoms $B$ do not contribute to the fluorescence signal because they
are already transferred to the $\left\vert F=1\right\rangle $ state in step
(III). The dynamics of photon scattering is described by the optical Bloch
equation, \cite{Metcalf} 
\begin{eqnarray}
\frac{d}{dt}\rho _{33} &=&-\Gamma \rho _{33}+\frac{i}{2}\Omega \left( \rho
_{13}-\rho _{31}\right) ,  \label{Bloch} \\
\frac{d}{dt}\rho _{13} &=&-\left( \frac{\Gamma }{2}+i\nu _{0}\right) \rho
_{13}+\frac{i}{2}\Omega \left( 2\rho _{33}-1\right) ,  \nonumber \\
\frac{d}{dt}\rho _{31} &=&-\left( \frac{\Gamma }{2}-i\nu _{0}\right) \rho
_{31}-\frac{i}{2}\Omega \left( 2\rho _{33}-1\right) ,  \nonumber
\end{eqnarray}%
where $\rho _{33}\left( t\right) =\left\vert c_{3}\left( t\right)
\right\vert ^{2}$ is the probability for the atom to be in the excited state 
$\left\vert 3\right\rangle $, $\left\vert 1\right\rangle $ represents state $%
\left\vert \uparrow \right\rangle $ for target atom $A$ at state $\left\vert
\uparrow \right\rangle $ and $\left\vert F=1\right\rangle $ for other cases, 
$\rho _{13}\left( t\right) =c_{1}\left( t\right) c_{3}^{\ast }\left(
t\right) e^{i\nu _{0}t}$, $\rho _{31}\left( t\right) =\rho _{13}^{\ast
}\left( t\right) $. The detuning of the laser $\nu _{0}$ is zero for target
atoms $A$ at state $\left\vert \uparrow \right\rangle $ and $\sim \nu $ for
all non-target atoms $B$ and part of the target atoms $A$ at hyperfine state 
$\left\vert 2\right\rangle $. $\Gamma =2\pi \times 6.07MHz$ is the decay
rate of the excited state $\left\vert 3\right\rangle $, $\Omega $ is the
Rabi frequency of the resonant laser that is related to the on-resonance
saturation parameter by $s_{0}=2\left\vert \Omega \right\vert ^{2}/\Gamma
^{2}$.

We numerically integrate the optical Bloch equation and calculate the number
of scattering photons 
\begin{equation}
n_{p}\left( t\right) =\Gamma \int_{0}^{t}\rho _{33}\left( t^{\prime }\right)
dt^{\prime }  \label{photon}
\end{equation}%
for both target and non-target atoms. In Fig. \ref{fig:error}a, we plot the
probability $\rho _{33}^{A}$ for target atoms $A$ at state $\left\vert
3\right\rangle $ with respect to time. We see $\rho _{33}^{A}$ increases
initially and reaches the saturation value $s_{0}/2\left( 1+s_{0}\right) $.
The number of scattering photons reaches $20$ in a short period, $1.3\mu s$
for atoms $A$ at $\left\vert \uparrow \right\rangle $ (Fig. \ref{fig:error}%
b), but it is only $10^{-5}$ for non-target atoms $B$ and target atoms $A$
at state $\left\vert 2\right\rangle $ (Fig. \ref{fig:error}(c)). Therefore
the impact of the resonant laser on the non-target atoms $B$ can be
neglected. In Fig. \ref{fig:error}(d) and (e), we see that for a wide range
of Rabi frequencies (laser intensities), the scattering photon number for
the non-target atoms $B$ is suppressed to undetectable levels, below $%
10^{-4} $.

Unlike the noise correlation method, the accuracy of our detection scheme
does not scale with the number of total atoms $N$, but is determined only by
manipulation errors in the above steps. We roughly estimate that $n$-spin
correlations can be probed with a cumulative error $<n\times 10^{-2}$, which
is sufficient to measure both long and short range spin correlation
functions. Our estimate takes into account possible experimental errors in
all four steps, and the fact that the same experimental procedure is
repeated many times to determine the probability $P_{l}$. The total failure
probability of the four step detection process is $p\left( n\right) \sim
n\times 10^{-3}$ for $n$ target atoms to give an incorrect measurement of
the target atom quantum state for determining the quantity $P_{l}$. Assuming
that the experimental procedure is repeated $\nu $ times, the total
probability for obtaining one incorrect measurement result is about $\nu
p\left( n\right) $. Since one incorrect measurement result leads to an error 
$\sim 1/\nu $ in $P_{l}$, the expectation for the error is $\nu p\left(
n\right) \times 1/\nu =p\left( n\right) $, which should be chosen to be $%
1/\nu $ to minimize the total error. In addition, the uncertainty in
measurements of probability $P_{l}$ itself in repeated experiments is also
about $1/\nu $. Taking into account of all these errors, we find that $n$%
-spin correlations can be probed with an error that scales as $Cp\left(
n\right) $, where $C\sim 3$ in our rough estimate. As a conservative
estimate, we take $C=10$ to give an overall error in measuring $n$-spin
correlation function $\xi \left\{ \alpha _{j_{k}},k=1,...,n\right\} $ to be
less than $10p\left( n\right) \sim n\times 10^{-2}$.

We note that the scheme requires repeated production of nearly the same
condensate and repeated measurements, two standard techniques which have
been realized in many experiments. We have proposed a powerful technique for
investigating strongly-correlated spin models in optical lattices and now
consider one of several possible applications.

\section{Spin Wave Dynamics:}

Our technique can be used to investigate time-dependence of correlation
functions. In the following, we show how our scheme can be used to engineer
and probe spin wave dynamics in a straightforward example, the Heisenberg $XX
$ model realized in optical lattices with a slightly different
implementation scheme than the one discussed in the previous section.
Consider a Mott insulator state with one boson per lattice site prepared in
the state $\left\vert 0\right\rangle \equiv \left\vert
F=1,m_{F}=-1\right\rangle $ in a three dimensional optical lattice. By
varying the trap parameters or with a Feshbach resonance, the interaction
between atoms can be tuned to the hard-core limit. With large optical
lattice depths in the $y$ and $z$ directions, the system becomes a series of
one dimensional tubes with dynamics described by the Bose-Hubbard
Hamiltonian: 
\begin{equation}
H_{s}=-\chi \left( t\right) \sum_{j}\left( a_{j}^{\dag
}a_{j+1}+a_{j+1}^{\dag }a_{j}\right) .  \label{Ham3}
\end{equation}%
This Bose-Hubbard \ model can be solved exactly. It offers a testbed for
spin wave dynamics by mapping the Hamiltonian (\ref{Ham3}) onto the $XX$
spin model, $H_{XXZ}(-2\chi ;\Delta =0)$, with the Holstein-Primakoff
transformation \cite{Auerbach}.

We now study the time dependent behavior of the $XX$ model using our
proposed scheme. In the Heisenberg picture, the time evolution of the
annihilation operator can be written as: $a_{j}\left( t\right)
=\sum_{j^{\prime}}a_{j^{\prime}}\left( 0\right)
i^{j^{\prime}-j}J_{j^{\prime}-j} \left( \alpha \right) $, where $%
J_{j^{\prime}-j}\left( \alpha \right) $ is the Bessel function of the
interaction parameter $\alpha(t) =2\int_{0}^{t}\chi \left( t^{\prime
}\right) dt^{\prime }$. To observe spin wave dynamics, we first flip the
spin at one site from $\uparrow $ to $\downarrow $, which, in the bosonic
degrees of freedom, corresponds to removing an atom at that site. Because of
the spin-spin interactions, initial ferromagnetic order gives way to a
re-orientation of spins at neighboring sites which propagates along the spin
chain in the form of spin waves. This corresponds to a time dependent
oscillation of atom number at each site. Therefore, spin wave dynamics can
be studied in one and two point spin correlation functions by detecting the
oscillation of the occupation probability at certain sites and the
density-density correlator between different sites, respectively.

Single atom removal at specific sites can be accomplished with the
assistance of focused lasers. With a combination of microwave radiation and
two focused lasers (propagating along $y\,$\ and $z$ directions
respectively), we can selectively transfer an atom at a certain site from
the state $\left\vert 0\right\rangle $ to the state $\left\vert
4\right\rangle \equiv \left\vert F=2,m_{F}=-2\right\rangle $. A laser
resonant with the transition $\left\vert 4\right\rangle \rightarrow
\left\vert 3\right\rangle \equiv \left\vert
5^{2}P_{3/2}:F=3,m=-3\right\rangle $ is then applied to remove an atom at
that site. Following an analysis similar to the one above, we see that the
impact on other atoms can be neglected. To observe fast dynamics of spin
wave propagation, we may adiabatically ramp down the optical lattice depth
(and therefore increase $\chi $) from the initial depth $V_{0}=50E_{R}$ to a
final depth $13E_{R}$, with a hold time, $t_{\text{hold}}$, to let the spin
wave propagate. Finally, the lattice depth is adiabatically ramped back up
to $V_{0}$ for measurement. The time dependence of the lattice depth in the
ramping down process is chosen to be 
\begin{equation}
V\left( t\right) =V_{0}/\left( 1+4\sqrt{2P_{\text{exe}}V_{0}/E_{R}}\omega
_{r}t\right) ,  \label{pot}
\end{equation}%
where $P_{\text{exe}}$ is the probability of making an excitation to higher
bands and $\omega _{R}=E_{R}/\hbar $. For $P_{\text{exe}}=4\times 10^{-4}$,
we find the interaction parameter to be $\alpha (t_{\text{hold}%
})=0.0146+0.0228\omega _{R}t_{\text{hold}}$, with the tunneling parameter: 
\begin{equation}
\chi \left( t\right) =\left( 4/\sqrt{\pi }\right) E_{R}^{1/4}V^{3/4}\left(
t\right) \exp \left( -2\sqrt{V\left( t\right) /E_{R}}\right) .  \label{tun}
\end{equation}


\begin{figure}[tbp]
\includegraphics[scale=0.4]{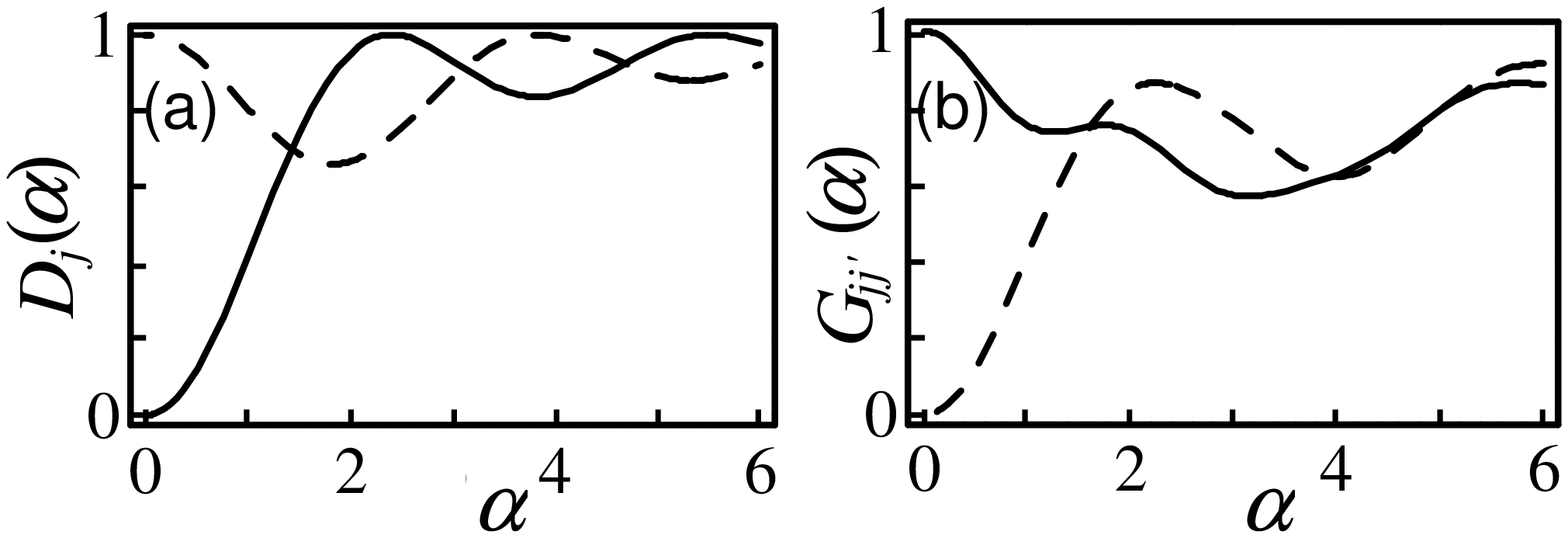}
\caption{Site occupation probability (a) and density-density correlation (b)
with respect to scaled spin interaction parameter $\protect\alpha \propto 
\text{time}$ for $N=30$ and: $j=0$ ($(a)$-solid line); $j=1$ ($(a)$-dashed
line); $j=0$, $j^{\prime }=3$ ($(b)$-dashed line); and $j=2$, $j^{\prime }=3$
($(b)$-solid line). The site index $j$ ranges from $-N$ to $N$. }
\label{fig:dynamics}
\end{figure}


Two physical quantities that can be measured in experiments are the single
atom occupation probability 
\begin{equation}
D_{j}\left( \alpha \right) =\left\langle \varphi \right\vert
a_{j}^{+}a_{j}\left\vert \varphi \right\rangle =\sum_{l\neq \kappa
}J_{l-j}^{2}\left( \alpha \right)  \label{ocpp}
\end{equation}%
at the site $j$, and the density-density correlator 
\begin{eqnarray}
&&G_{jj^{\prime }}\left( \alpha \right) =\left\langle \varphi \right\vert
a_{j}^{+}a_{j}a_{j^{\prime }}^{+}a_{j^{\prime }}\left\vert \varphi
\right\rangle  \label{dencor} \\
&=&\sum\limits_{l\neq \kappa ,\gamma \neq \kappa }J_{l-j}^{2}\left( \alpha
\right) J_{\gamma -j^{\prime }}^{2}\left( \alpha \right) -\sum\limits_{l\neq
\kappa }(J_{l-j}^{2}\left( \alpha \right) J_{l-j^{\prime }}^{2}\left( \alpha
\right)  \nonumber \\
&&+J_{l-j}\left( \alpha \right) J_{\kappa -j}\left( \alpha \right) J_{\kappa
-j^{\prime }}\left( \alpha \right) J_{l-j^{\prime }}\left( \alpha \right) ) 
\nonumber
\end{eqnarray}%
between sites $j$ and $j^{\prime }$, where $\varphi $ is the initial
wavefunction with one removed atom at site $\kappa $. The former is related
to the local transverse magnetization through 
\begin{equation}
\left\langle \varphi \right\vert s_{j}^{z}\left( \alpha \right) \left\vert
\varphi \right\rangle =D_{j}\left( \alpha \right) -1/2,  \label{mag}
\end{equation}%
and the latter is related to the spin-spin correlator via 
\begin{eqnarray}
G_{jj^{\prime }}\left( \alpha \right) &=&\left\langle \varphi \right\vert
s_{j}^{z}\left( \alpha \right) s_{j^{\prime }}^{z}\left( \alpha \right)
\left\vert \varphi \right\rangle  \label{spincor2} \\
&&+\left( D_{j}\left( \alpha \right) +D_{j^{\prime }}\left( \alpha \right)
\right) /2+1/4.  \nonumber
\end{eqnarray}

In Fig. 4, we plot $D_{j}\left( \alpha \right) $ and $G_{jj^{\prime }}\left(
\alpha \right) $ with respect to the interaction parameters $\alpha $ (which
scales linearly with holding time). Here the initial empty site at $j=0$ is
located at the center of the spin lattice. We see different oscillation
behavior at different sites. Initially, all sites are occupied except $j=0$,
i.e., $D_{j\neq 0}\left( 0\right) =1$, $D_{j=0}\left( 0\right) =0$
(Fig.4(a)). The initial spin-spin correlation $G_{jj^{\prime }}\left(
0\right) $ between different sites is zero if either $j$ or $j^{\prime }$ is
zero, and one if both of them are nonzero (Fig.4(b)). As $\alpha $
increases, atoms start to tunnel between neighboring sites and the spin wave
propagates along the one dimensional optical lattice, which is clearly
indicated by the increase (decrease) of the site occupation at $j=0$ ($j\neq
0$). In Fig.4, the oscillation of $D_{j}\left( \alpha \right) $ and $%
G_{jj^{\prime }}\left( \alpha \right) $ in a long time period and the decay
of the oscillation amplitudes originate from the finite size of the spin
lattice, which yields the reflection of the spin waves at the boundaries.

To probe the single site occupation probability $D_{j}\left( \alpha \right) $%
, we use two focused lasers and one microwave pulse to transfer the atom at
site $j$ to $\left\vert 4\right\rangle $. A laser resonant with the
transition $\left\vert 4\right\rangle \rightarrow \left\vert 3\right\rangle $
is again applied to detect the probability to have an atom at $\left\vert
4\right\rangle $, which is exactly the occupation probability $D_{j}\left(
\alpha \right) $. To detect $G_{jj^{\prime }}\left( \alpha \right) $, we
transfer atoms at both sites $j$ and $j^{\prime }$ to the state $\left\vert
4\right\rangle $ and use the same resonant laser to detect the joint
probability for atoms at $\left\vert 4\right\rangle $. The fluorescence
signal has three levels, which correspond to both atoms $G_{jj^{\prime
}}\left( \alpha \right) $, one atom $D_{j}\left( t\right) +D_{j^{\prime
}}\left( t\right) $, and no atoms at state $\left\vert 4\right\rangle $. A
combination of these measurement results gives the spin-spin correlator $%
\left\langle \varphi \right\vert s_{j}^{z}\left( \alpha \right) s_{j^{\prime
}}^{z}\left( \alpha \right) \left\vert \varphi \right\rangle $.

\section{Conclusion}

We find that a relation between general spin correlation functions and
observable state occupation probabilities in optical lattices allows for
quantitative measurements of a variety of spin correlators with the help of
local probes, specifically focused lasers and microwave pulsing. Our
proposal includes a realistic and practical quantitative analysis suggesting
that $n$-spin correlations can be probed with an error $<n\times 10^{-2}$,
which is sufficient to measure both long and short range spin correlation
functions. Our work establishes a practical and workable method for
detecting $n$-spin correlations for cold atoms in one, two or three
dimensional optical lattices. Applications to a broad class of spin physics
including topological phases of matter \cite{Kitaev,Zhang1} realized in
spin-optical lattices are also possible with our proposed technique.

This work is supported by ARO-DTO, ARO-LPS, and LPS-NSA.

\end{document}